\documentclass[aps,pre,twocolumn,groupedaddress]{revtex4-2}
\usepackage{amsmath,amsfonts,mathtools,amssymb}
\usepackage[colorlinks=true,linkcolor=blue,citecolor=red,bookmarksnumbered=true]{hyperref}
\usepackage{times,mathpazo,charter,graphicx,bm,enumitem}
\usepackage{color}
\usepackage{ulem}
\definecolor{deeppurple}{rgb}{0.5,0,0.5}

\newcommand{\orcid}[1]{\href{https://orcid.org/#1}{\includegraphics[width=10pt]{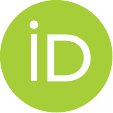}}}
\graphicspath{{figs/}}

\makeatletter
\def\overl@ss#1#2{\vcenter{\offinterlineskip
      \ialign{$\m@th#1\hfil##\hfil$\crcr#2\crcr<\crcr } }}
\def\overgr@at#1#2{\vcenter{\offinterlineskip
      \ialign{$\m@th#1\hfil##\hfil$\crcr#2\crcr>\crcr } }}
\def\gl{\mathrel{\mathpalette\overl@ss>}}
\def\lg{\mathrel{\mathpalette\overgr@at<}}
\makeatother

\graphicspath{{figs/}}

\def\d{{\mathrm{d}}}
\def\L{{\mathcal{L}}}

\def\dn{\mathop{\rm dn}\nolimits}
\def\sn{\mathop{\rm sn}\nolimits}
\def\cn{\mathop{\rm cn}\nolimits}
\def\am{\mathop{\rm am}\nolimits}
\def\sech{\mathop{\rm sech}\nolimits}
\def\diag{\mathop{\rm diag}\nolimits}

\def\Real{\mathbb{R}}
\def\Complex{\mathbb{C}}
\def\Natural{\mathbb{N}}
\def\N{\mathbb{N}}
\def\Re{\mathop{\rm Re}\nolimits}
\def\Im{\mathop{\rm Im}\nolimits}
\def\tr{\mathop{\rm tr}\nolimits}

\def\d{\mathrm{d}}
\def\e{\mathop{\rm e}\nolimits}
\def\@#1{{\mathbf{#1}}}
\def\_#1{{\mathrm{#1}}}
\def\q{q}
\def\D{\Delta}
\def\half{{\textstyle\frac12}}

\def\R{\mathbb{R}}
\def\Z{\mathbb{Z}}
\let\~=\tilde

\def\be{\begin{equation}}
\def\ee{\end{equation}}
\def\bse{\begin{subequations}}
\def\ese{\end{subequations}}
\def\eqref#1{(\ref{#1})}

\def\note[#1]{\marginpar{\color{red}[#1]}}

\long\def\paragraph#1{\par\smallskip\textbf{#1.}}

\newcommand\GE[1]{{\color{magenta} \it GE: #1}}

\allowdisplaybreaks

\def\reftitle#1{``#1''}

\def\titletext{Breather  gas fission from elliptic potentials in self-focusing media}

\begin{document}
\title{\Large\titletext}
\author{Gino Biondini\orcid{0000-0003-3835-1343}$^{1,2}$, 
Gennady A. El\orcid{0000-0003-1962-5388}$^3$
Xu-Dan Luo\orcid{0000-0002-3833-5530}$^4$, Jeffrey Oregero\orcid{0000-0003-1576-8867}$^5$, and Alexander Tovbis$^6$\\[-1.6ex]~}
\affiliation{$^1$ Department of Mathematics, State University of New York at Buffalo, Buffalo, New York 14260, USA}
\affiliation{$^2$ Department of Physics, State University of New York at Buffalo, Buffalo, New York 14260, USA}
\affiliation{$^3$ Department of Mathematics, Physics and Electrical Engineering, Northumbria University, Newcastle upon Tyne, NE1 8ST, United Kingdom}
\affiliation{$^4$ Academy of Mathematics and Systems Science, Chinese Academy of Sciences, Beijing 100190, China}
\affiliation{$^5$ Department of Mathematics, University of Kansas, Lawrence, KS 66045, USA}
\affiliation{$^6$ Department of Mathematics, University of Central Florida, Orlando, Florida 32816, USA}
\date{\today}

\begin{abstract}
\noindent
We present an analytical model of integrable turbulence in the focusing nonlinear Schrödinger (fNLS) equation, 
generated by a one-parameter family of finite-band elliptic potentials in the semiclassical limit.
We show that the spectrum of these potentials exhibits a thermodynamic band/gap scaling compatible with that of soliton and breather gases depending on the value of the elliptic parameter $m$ of the potential. 
We then demonstrate that, upon augmenting the potential by a small random noise 
(which is inevitably present in real physical systems), 
the solution of the fNLS equation evolves into a fully randomized, spatially homogeneous breather gas, 
a phenomenon we call breather gas fission. 
We show that the statistical properties of the breather gas at large times are determined by the spectral density of states generated by the unperturbed initial potential. 
We analytically compute the kurtosis of the breather gas as a function of the elliptic parameter $m$, 
and we show that it is greater than 2 for all non-zero $m$, 
implying non-Gaussian statistics.
Finally, we verify the theoretical predictions by comparison with direct numerical simulations of the fNLS equation. 
These results establish a link between semiclassical limits of integrable systems and the statistical characterization of their soliton and breather gases. \end{abstract}
\maketitle


\textbf{Introduction.}
The focusing nonlinear Schr\"odinger (fNLS) equation is a ubiquitous model describing nonlinear wave propagation  arising in a variety of physical settings, including
deep water waves \cite{AS1979}, optics \cite{Agrawal2007}, plasmas \cite{IR2000}, and Bose-Einstein condensates (BECs) \cite{PS2003}. 
The fNLS equation is also a completely integrable infinite-dimensional Hamiltonian system \cite{AS1981,NMPZ1984,FT1987,BBEIM1994,GesztesyHolden,APT2004},
endowed with a deep mathematical structure, including the existence of 
infinite families of exact solutions with both zero and non-zero background
describing the elastic interactions of $N$ solitons.
In addition, its initial value problem can in some cases be solved by the inverse scattering transform (IST) \cite{GGKM,Lax, ZS1972,ZS1973,ItsKotlyarov,MaAblowitz,ForestLee,mclaughlinoverman,BK2014}.
The formulation of the IST is based on the representation of the nonlinear evolution equation as the compatibility condition of two linear equations
called a Lax pair.
The first half of the Lax pair of the fNLS equation, namely the Zakharov-Shabat (ZS) scattering problem,  
is equivalent to an eigenvalue problem for a non-self-adjoint one-dimensional Dirac operator.

While the classical theory and applications of the fNLS equation are mostly concerned with the description of regular, deterministic  wave structures, the inherent statistical nature of some physical wave phenomena in focusing media (e.g., rogue wave emergence) calls for the study of stochastic fNLS solutions,  characterized in terms of the probability density function, correlation function etc. 
Establishing a connection between the IST spectra of random fNLS solutions and their  statistical properties in physical space represents a challenging problem, which has recently been formulated in the context of {\it integrable turbulence}  \cite{Zakharov2009,Congy2024} --- the general theoretical framework for the description of a broad spectrum of stochastic wave phenomena in physical systems modeled by integrable equations. 
A particular  type of integrable turbulence, termed \textit{soliton gas} (SG),  
has recently attracted considerable attention \cite{El2021,PRE_persp} due to its appearance in many physical systems including water waves \cite{costa2014,redor2019,suret2020,mordant2D}, 
nonlinear optics \cite{PRE1997v55mitschke,marcucci2019,PhysD1996v96} 
and BECs \cite{Bouchoule}.
Notably, SG dynamics has been shown to  underpin some fundamental physical phenomena such as spontaneous modulational instability \cite{PRL2019v123p234102} 
and the rogue wave emergence \cite{PRE98p042210} 
in  focusing media.  
 
The concept of a SG was introduced in \cite{Zakharov71} as an infinite collection of randomly distributed solitons with small spatial density and with a certain amplitude distribution. 
Soliton interactions, accompanied by well-defined phase shifts, result in a modification of the effective velocity of a ``tracer'' soliton in a gas over large propagation distances, enabling  an approximate  description of the emergent, large-scale hydrodynamics or kinetics of a weakly non-uniform/non-equilibrium SG. 
The kinetic description of a SG was generalized in \cite{El2003}  
to a dense SG using finite-gap theory, 
and a general phenomenological construction of SG kinetic equation for a broad class of integrable systems  was proposed in \cite{ElandKam}.
A systematic spectral theory of fNLS SGs  was developed in \cite{ElTovbis}, 
where it was also extended to the case of solutions on a non-zero background,
i.e., a \textit{breather gas} (BG). 
For recent advances on the spectral theory of soliton and breather gases and its relation to the generalised hydrodynamics of integrable many-body classical and quantum systems,
see~\cite{PRE_persp}.

While BGs represent a natural generalization of SGs, 
possible mechanisms for their generation have remained largely unexplored. 
In this Letter we present an analytically tractable model of BG generation via fission, 
based on the semiclassical limit of the fNLS equation with initial data in the form of a  periodic elliptic ``$\dn$'' potential with elliptic parameter $m \in [0,1]$, augmented by a small random noise. 
It was shown in \cite{BLOT2023}
that the ZS spectrum of such elliptic potentials can be characterized analytically. 
Here we show that, in the semiclassical limits, these potentials are compatible with the so-called thermodynamic spectral scaling \cite{ElTovbis} and give rise to a bound state BG. 
The addition of a small noise to the initial condition(IC) facilitates the ``phase mixing'' 
of the finite-gap semiclasical $\dn$ potential, giving rise, in the long-time limit,  
to a spatially uniform and statistically stationary integrable turbulence, 
associated with a BG. 
The fundamental property of long-time ``thermalization'', or relaxation to a statistically stationary state, in an integrable system was established numerically in several scenarios of the evolution of random waves in the fNLS equation 
\cite{agafonstev2015,agafontsev2016,suret2021}.   
Here we extend this result to the qualitatively new framework of the BG fission, 
which encompasses a broad range of scenarios of transition to stationary integrable turbulence depending on the value of $m$ in the semiclassical elliptic potential: 
from the nonlinear development of the spontaneous modulational instability for $m \to 0$ to the rarefied soliton gas fission for $m \to 1$.
When the noise is sufficiently small, the ZS spectrum remains essentially unchanged from that of the semiclassical elliptic potential.  
The isospectrality of the BG fission enables one to take advantage of the results of~\cite{Congy2024} and evaluate statistical measures of the BG, such as the mean intensity and kurtosis, in terms of the spectral density of states of the initial elliptic potential. 
In \cite{Congy2024},
it was predicted that the kurtosis doubles in the long-time fNLS fission of the so-called partially coherent waves into a SG.
Here we show analytically that this result generalizes to the BG fission from semiclassical elliptic potentials,
and we confirm our results by comparison with direct numerical simulations of the time evolution of the noise-augmented potential. 

%
An important ingredient of our construction is the interpretation of a BG as a ``composite SG'' comprising of two spectrally distinct components: a regular SG  and a {\it soliton condensate}, defined as a critically dense SG \cite{ElTovbis} and
describing the modulationally unstable background in a BG. This composite SG   
provides a natural extension of the  
 ``pure'' fNLS soliton condensate framework used in  \cite{PRL2019v123p234102, agaf_dn} to model the development of spontaneous (noise-induced) modulational instability of fNLS plane wave and genus one elliptic solutions.


\paragraph{NLS and semiclassical elliptic potentials}
The cubic fNLS equation is, in normalized and dimensionless form and in the semiclassical scaling,
\vspace*{-0.4ex}
\begin{gather}
\label{e:nls}
i\epsilon \q_t + \epsilon^2\q_{xx} + 2\,|\q|^2 \q = 0\,,
\end{gather}
where subscripts $x$ and~$t$ denote partial differentiation, 
$\q(x,t)$ describes a complex-valued envelope of oscillations,
and the physical meaning of the variables $x$ and~$t$ varies depending on the physical context.
(E.g., in nonlinear fiber optics, $x$ is a retarded time and $t$ is the propagation distance through the medium.)
Equation~\eqref{e:nls} is the compatibility condition $\@v_{xt} = \@v_{tx}$ of its Lax pair, namely the overdetermined linear system
\vspace*{-0.6ex}
\be
\epsilon\@v_x = (-iz\sigma_3 + Q)\,\@v\,,\quad 
\epsilon\@v_t = P\,\@v\,,
\label{e:LP}
\ee
for $\@v(x,t,z) = (v_1,v_2)^T$, with 
$P = -2iz^2\sigma_3 + i(|q|^2+Q_x)\,\sigma_3 -2zQ$, 
where 
$\sigma_3 = \diag(1,-1)$ is the third Pauli matrix, and
\vspace*{-1.2ex}
\be
Q(x,t) = \begin{pmatrix} 0 &\q \\ - \q^* & 0 \end{pmatrix}
\vspace*{-1ex}
\ee
the asterisk denoting complex conjugate.
The first half of Eq.~\eqref{e:LP} is the ZS scattering problem,
and $\q$ and $z$ are referred to respectively as the potential and the scattering parameter, or eigenvalue,
since the first half of Eq.~\eqref{e:LP} can be rewritten as the eigenvalue problem
$\L\@v = z\@v$,
where $\L := i\sigma_3(\epsilon\partial_x - Q)$ is a one-dimensional Dirac operator.

\begin{figure}[t!]
\smallskip
\centerline{\hglue-1em\includegraphics[trim=35 0 50 30,clip,height=0.205\textwidth]{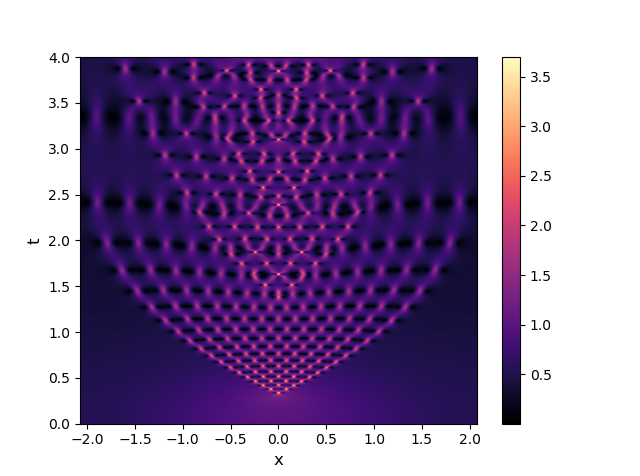}\kern0em
\includegraphics[trim=35 0 50 30,clip,height=0.205\textwidth]{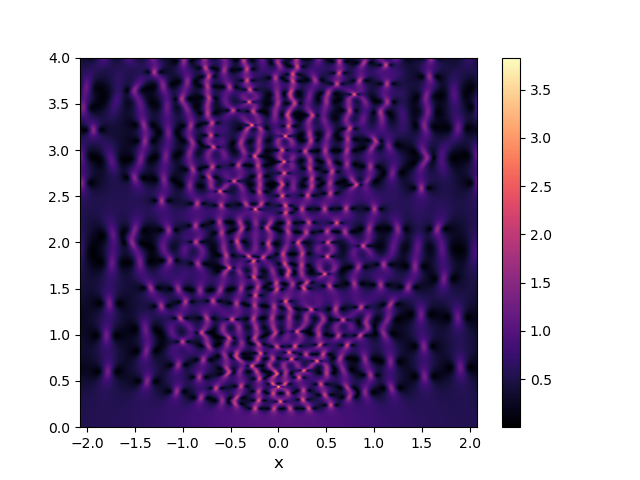}\kern-2em}
\kern-\smallskipamount
\caption{Density plots of the time evolution of the potential in Eq.~\eqref{e:ellipticpotential} according~to fNLS equation \eqref{e:nls} with $\epsilon=1/20$ and:
(a)~$m=0.7$ without noise;
(b)~$m=0.7$ with noise (see text for details).
In both cases, the spatial window coincides with the period of the IC.
}
\label{f:nlsplots}
\end{figure}

In \cite{BLOT2023} we characterized the Lax spectrum $\Sigma(\L)$ \cite{Laxspecdef} 
of the above ZS problem with the potential 
\vspace*{-0.6ex}
\be
\q(x,0) = \dn(x;m)\,, \quad x\in\R\,,
\label{e:ellipticpotential}
\ee
where $\dn(x;m)$ is a Jacobi elliptic function~\cite{ByrdFriedman,NIST} 
and $m\in(0,1)$ is the elliptic parameter.
Note the spatial period is $2L = 2K_m$, where $K_m=K(m)$ is the complete elliptic integral of first kind. 
Two limiting cases are $m=0,1$.
The case $m=0$ reduces to the constant potential $\q(x,0)\equiv1$, which is solved trivially, and for which $\Sigma(\L) = \Real \cup [-i,i]$.
Conversely, as $m\to1$, $\q(x,0) \to \sech x$ and $K_m\to\infty$.
This problem was studied in~\cite{SatsumaYajima}, where 
it was shown that $\Sigma(\L)$ comprises the real $z$-axis plus a set of purely imaginary discrete eigenvalues
uniformly distributed in the interval $(-i,i)$.
Moreover, it was shown that $\sech x$ is a reflectionless potential 
(giving rise to pure soliton solutions) if and only if $\epsilon = 1/N$, with $N\in\Natural$. 
This Letter connects these two limiting cases by 
characterizing the class of periodic potentials~\eqref{e:ellipticpotential} for all $0<m<1$. 
Figure \ref{f:nlsplots} shows for $\epsilon=1/20$ and $m=0.7$ the time evolution of the dn potential according to fNLS without noise (left), and with Gaussian noise (right).

\paragraph{Band-gap structure}
The structure of the Lax spectrum can be studied via Bloch-Floquet theory using the monodromy matrix $M(z)=\Phi^{-1}(x,z)\Phi(x+2L,z)$, 
where $\Phi(x,z)$ is a fundamental matrix solution of the scattering problem, and $2L$ is the minimal period of the potential \cite{omit_time_dependence}.
In turn, the Floquet discriminant is defined by $\D(z)=\tr M(z)/2$. Bounded solutions of the scattering problem then exist for $z\in\mathbb{C}$ such that $\Im \D(z) = 0$ and 
$-1\le \Re\D(z)\le 1$. Importantly, $\D(z)$ is an entire function of $z$ that is also Schwarz-symmetric: $\D(z^*) = \D^*(z)$. As a result, it is sufficient to consider the upper half plane, $\Im z\ge0$. Since $\L$ is non-self-adjoint, $\Sigma(\L)$ is not confined to the real $z$-axis. 
Nonetheless, the zero-level curves of $\Im\D(z)$ define a countable set of analytic arcs $\Gamma_n$. Along each arc, the requirement $-1\le\Re\D(z)\le1$ then defines a spectral band. With these definitions, one can talk about bands as in a self-adjoint problem, the difference being that in the non-self-adjoint case the bands are not restricted to lie along the real $z$-axis, but lie instead along the~$\Gamma_n$.

The endpoints of the bands are periodic and antiperiodic eigenvalues of $\L$, namely, the values~$z$ for which $\D(z) = \pm1$, which give rise to periodic and antiperiodic eigenfunctions, respectively. 
Moreover, the real $z$-axis is an infinitely long band, and 
any band that intersects the real $z$-axis transversally is called a ``spine''. 
Any potential whose spectrum is comprised of a finite number of bands is called a ''finite-band'' potential \cite{BBEIM1994,GesztesyHolden,GW1998}.
These special potentials correspond to the finite-genus solutions of all equations in the Ablowitz-Kaup-Newell-Segur (AKNS) hierarchy. 

\begin{figure}[t!]
\kern-\medskipamount
\centerline{~~\includegraphics[width=0.215\textwidth]{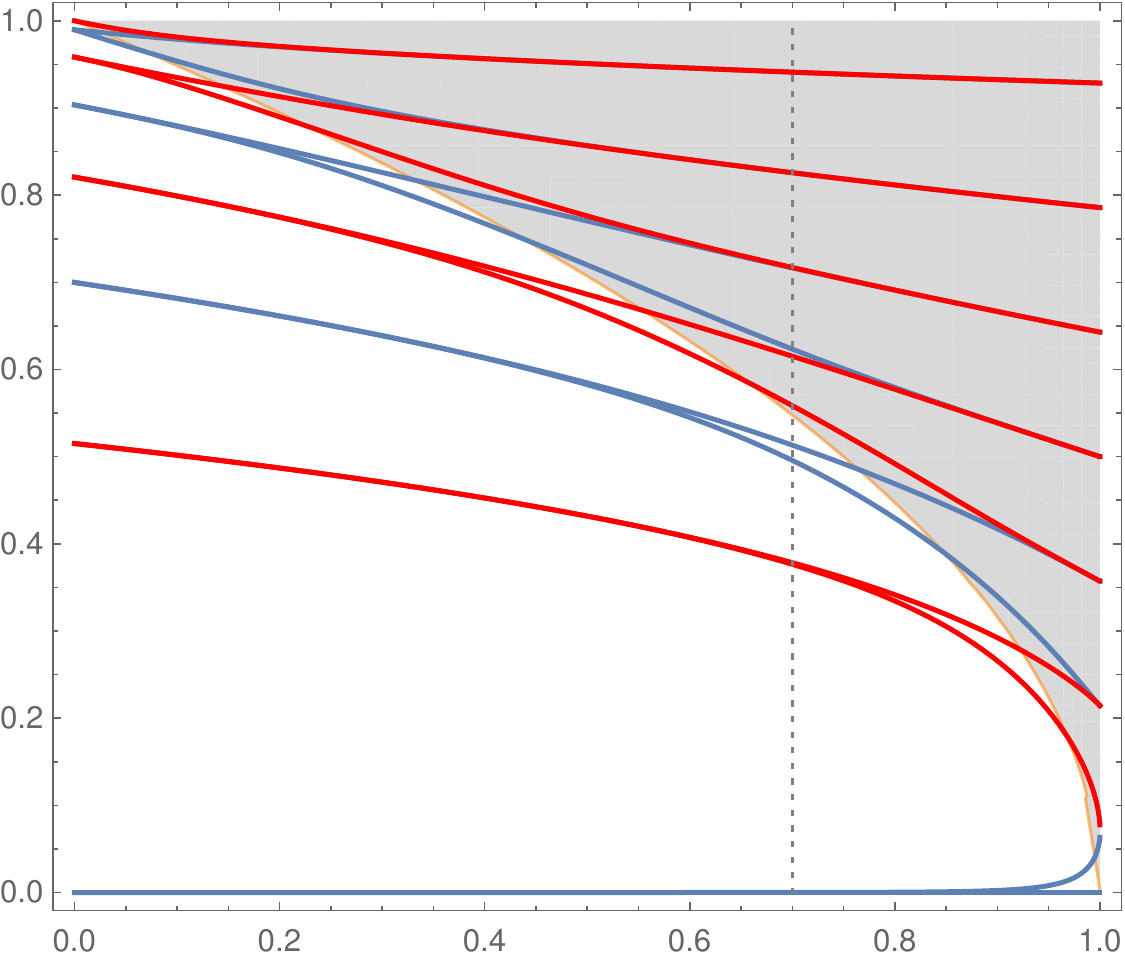}
\lower0.6ex\hbox{\includegraphics[width=0.2905\textwidth]{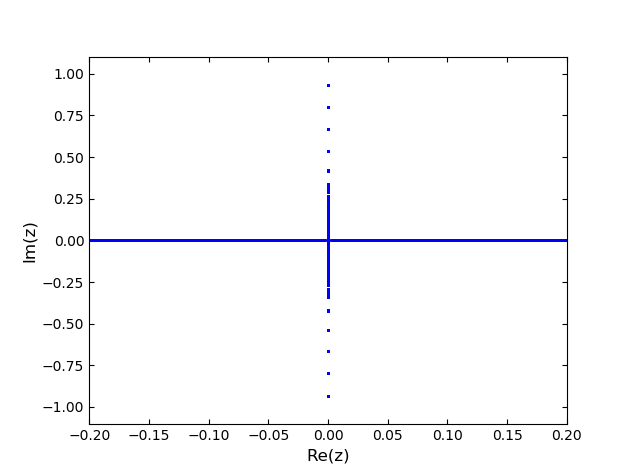}}\kern-1em}
\caption{%
Left: Periodic (red) and antiperiodic (blue) eigenvalues (vertical axis) as a function of the elliptic parameter~$m$ (horizontal axis) for $\epsilon = 1/7$.
For every fixed $m\in(0,1)$, as $\epsilon\to0$, the number of bands grows like $1/\epsilon$.
In the gray range, corresponding to $\Gamma_g$, the band widths decrease exponentially in $\epsilon$.
Conversely, in the white range, corresponding to $\Gamma_o$, the band widths decrease algebraically.
Right: The spectrum of the dn potential~\eqref{e:ellipticpotential} in the complex $z$ plane with $m=0.9$ for $\epsilon = 1/7$.
}
\label{f:spectrum}
\end{figure}

\paragraph{Elliptic finite-band potentials}
In \cite{BLOT2023} it was proved that for the dn potential \eqref{e:ellipticpotential} and $\epsilon=1/N$, $N\in\N$, the Lax spectrum of the ZS scattering problem comprises $2N$ Schwarz-symmetric bands along the interval $(-i,i)$, and produces a genus $2N-1$ solution of fNLS \cite{supplement}. Figure~\ref{f:spectrum} (left) illustrates the dependence of the spectrum on the elliptic parameter~$m$ (horizontal axis), in particular, the periodic (red curves) and antiperiodic (blue curves) eigenvalues of $\L$ along the imaginary $z$-axis (vertical axis) 
for $\epsilon = 1/7$, computed using finite truncations of Eqs.~\eqref{e:B_oinftypm}. Notice all gaps are closed when $m=0$ and they open as $m>0$ and remain open for all $m\in(0,1)$.
In the singular limit $m\to1$, the band widths tend to zero, and the periodic and antiperiodic eigenvalues ``collide'' to form the point spectrum of~$\L$ on the line. The corresponding spectrum in the complex $z$ plane for $m=0.9$ is shown in Fig.~\ref{f:spectrum} (right).

\paragraph{Semiclassical limit and breather gas}
Even though $\L$ is not self-adjoint, its spectrum in the limit $\epsilon\to 0$ can be effectively characterized using WKB methods~\cite{BO2020,BOT}, 
which allows one to obtain precise asymptotic estimates for the location of band edges, band widths, and gap widths~\cite{supplement}. 
In turn, these calculations show that the semiclassical limit of the spectrum of the elliptic potential~\eqref{e:ellipticpotential} 
is compatible with the thermodynamic spectral scaling of a SG or a BG~\cite{ElTovbis,Wang2022}. 
All bands lie along a finite Schwarz-symmetric arc $\Gamma = (-iq_\mathrm{max},iq_\mathrm{max})$ 
in the complex $z$-plane. 
As $\epsilon\to0$, however,
the behavior of bands and gaps in the range $\Gamma_o = (0,iq_\mathrm{min})$
differs from those in $\Gamma_g = (iq_\mathrm{min},iq_\mathrm{max})$.
Inside $\Gamma_g$, 
band widths 
decay exponentially in $\epsilon$, the ratio of band width to gap width tends to zero,
and the bands accumulate according to some limiting density $\varphi(z)$~\cite{supplement}.
Conversely, inside $\Gamma_o$, it is the relative band gaps that tend to zero,
giving rise to a continuous band in the limit.
Thus, exponentially small bands outside the finite band give rise to an ``effective soliton ensemble'' on an ``effective nonzero background'', i.e., a generalized BG. 
As $m\to1$, the range $\Gamma_o$ tends to zero, and one gets a pure SG.
Conversely, as $m\to0$, it is the range $\Gamma_g$ that tends to zero, and one obtains a pure condensate.
We therefore have that, in the semiclassical limit, 
\textit{the potential~\eqref{e:ellipticpotential} gives rise to a composite SG}.
Next we characterize this gas in detail.


We can derive the two key quantities in the spectral theory of soliton and breather gases, namely, the density of bands, and the normalized logarithmic band width. 
In particular, the density of bands $\varphi(z)$ is \cite{supplement}
\be
\label{e:varphi}
\varphi(z) = \frac{2|z|}{\pi}\int_0^{x_o(z)}[\dn^2(x;m)-|z|^2]^{-1/2}\,\d x \,,
\ee
where $x_o(z) = K_m$ for $z\in(0,iq_\mathrm{min})$ 
and $x_o(z)>0$ is a simple turning point [at which $\dn(x;m) = |z|$]
for $z\in(iq_\mathrm{min},i)$
\cite{integralone}.
Following \cite{ElTovbis,El2021,Wang2022}, we obtain the 
density of states (DOS) $f(z)$ as \cite{reducedDOS,supplement}
$f(z) = \varphi(z)/(2K_m)$.
As $m\to0$, $\Gamma_g$ collapses, and $f(\eta)$ tends to the Weyl distribution of a soliton condensate \cite{supplement}.
Conversely, as $m\to1$, $L\to\infty$, $\Gamma_o$ collapses, and $f(\eta)$ tends to zero, as appropriate for a rarified SG. 
The interpretation of a BG as a composite soliton gas is also supported by the results of \cite{agafontsev_etal_SAPM_2024}, where basic fNLS breathers were generated numerically 
with high accuracy by appropriately configuring $N$-soliton fNLS solutions with large $N$.

\paragraph{Kurtosis}
The DOS allows one to obtain the following expression for the kurtosis $\kappa$. 
In~\cite{Congy2024}, the following general expression for the kurtosis $\kappa$ of spatially uniform bound state integrable turbulence was obtained:
\vspace*{-0.8ex}
\be 
\label{kurtosis}
\kappa = 
  \frac{\langle |q|^4 \rangle }{\langle |q|^2\rangle^2} = \frac{2\overline{\eta^3}}{3\overline{\eta}^2}\,,
\ee
where the angle brackets denote ensemble average,
$\eta = \Im z$,
and the overbar denotes average over the DOS: 
\vspace*{-0.4ex}
\be
\overline{\eta^k} = \int_{\gamma} \eta^k f(\eta)\,\d \eta
  = (-1)^{\frac{k+1}{2}}\frac{c_k}L \int_0^L q^{k+1}(x,0)\,\d x\,
\ee
are the moments of the  DOS $f(\eta)$,
where $\gamma$ is the spectral support of the SG, 
and where the last equality follows from \cite{Wang2022},
with $c_1= \frac14$ and $c_3= \frac3{16}$.
Thus,
\vspace*{-0.6ex}
\bse
\begin{gather}
	\kappa = 2\kappa_0,
\label{e:kdouble}
\\
\noalign{\noindent 
where $\kappa_0$ is the normalized fourth moment of the IC:} 
 \kappa_0 = L \frac{\int_0^{L} |q(x,0)|^4\,\d x }{ \big( \int_0^{L} |q(x,0)|^2\,\d x\big) ^2}\,.
\label{e:k0}
\end{gather}
\ese
This result is consistent with the kurtosis doubling obtained in \cite{Congy2024} for the
SG fission from partially coherent waves, 
with a crucial difference: 
here $\kappa_0$ is computed {\it from a purely deterministic IC}. 
Moreover, the above results imply that \cite{supplement} 
\textit{any periodic SG or BG of the fNLS equation generated by a deterministic 
real and even single-lobe potential $q(x,0)$ has kurtosis $\kappa\ge 2$,
with $\kappa=2$ if and only if $q(x,0)$ is constant.} 
Recall that $\kappa >2$ implies heavy-tailed non-Gaussian statistics, 
which is an indication of the possible presence of rogue waves 
\cite{physrep2013v528p47,walczak2015,revphys2020v5p100037}.
In particular, for the potential~\eqref{e:ellipticpotential},
straightforward calculations yield
\vspace*{-0.6ex}
\be
\kappa_0 = K_m[2(2-m)E_m - (1-m)K_m]/(3E_m^2)\,,
\label{e:kappa0}
\ee
\hbox{with $E_m$ the complete elliptic integral of the second kind.}

\paragraph{Temporal dynamics and numerical validation}
Next we discuss the numerically computed time evolution of the fNLS solutions produced by the potential~\eqref{e:ellipticpotential}.
Figure~\ref{f:nlsplots} demonstrates how the addition of a small amount of 
zero-mean white Gaussian noise (with standard deviation $\sigma = 10^{-2}$) to the IC
effectively randomizes the solution.
Figure~3(left) shows the temporal evolution of the normalized fourth moment $|q|^4/(|q|^2)^2$ of the resulting fNLS solutions as well as their ensemble average (i.e., the kurtosis), 
clearly demonstrating the establishment of a statistically stationary integrable turbulence state.
Finally Fig.~3(right) shows how the dependence of the kurtosis on the elliptic parameter~$m$
matches the theoretical prediction from Eqs.~(\ref{kurtosis}--\ref{e:kappa0}),
providing a strong validation of the analytical predictions. 

\begin{figure}[t!]
\kern-\bigskipamount
\centerline{\kern-1em%
\includegraphics[width=0.275\textwidth]{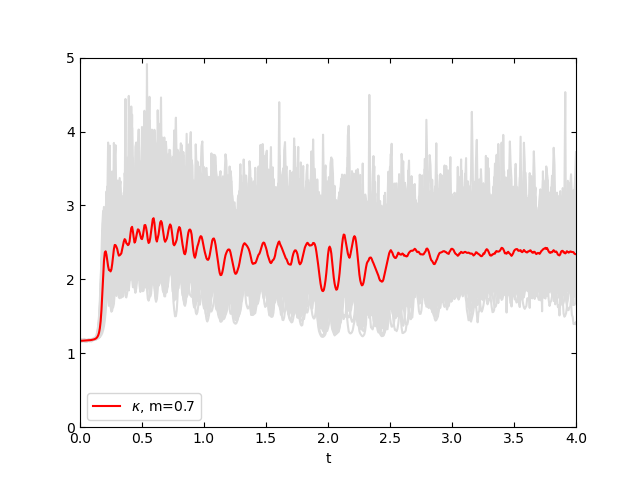}\kern-0.8em%
\raise0.6ex\hbox{\includegraphics[width=0.235\textwidth]{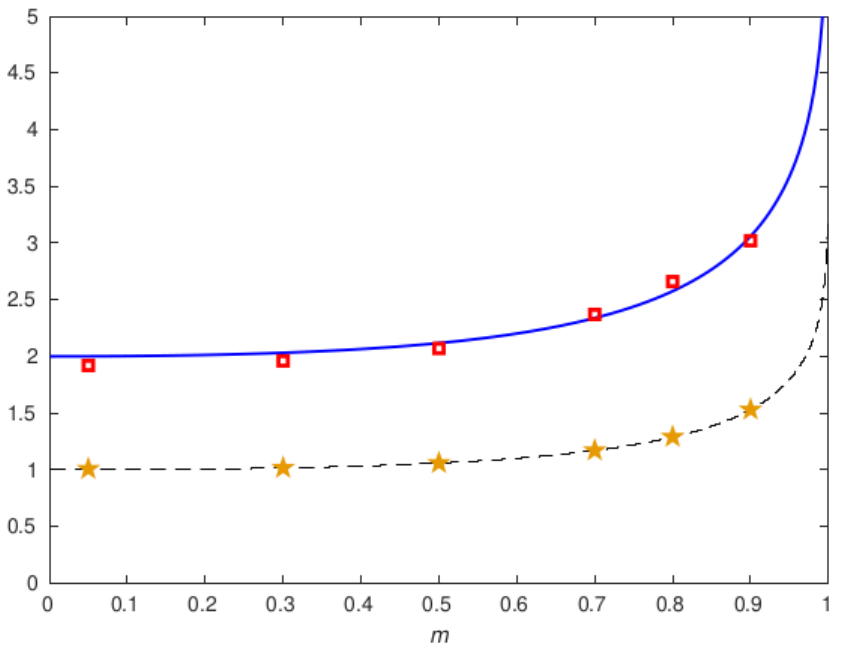}}}
\caption{
Left: The averaged temporal evolution of the kurtosis from an ensemble of noise realizations
(red line).  The gray region shows the envelope of the normalized fourth moment of the solution from all individual noise realizations.
Right: The kurtosis of a fully randomized gas as a function of the elliptic parameter~$m$.
Blue curve: theoretical prediction from Eqs.~\eqref{e:kdouble} and~\eqref{e:kappa0}.
Dashed black curve: $\kappa_0$ from Eq.~\eqref{e:kappa0}.
Red squares: Numerically computed asymptotic value of kurtosis.
Yellow stars: Ensemble average of the numerically evaluated integral from Eq.~\eqref{e:k0} for ICs consisting of the dn potential plus noise.
}
\label{f:kurtosis}
\end{figure}

\paragraph{Concluding remarks}
In summary, we have presented an analytically tractable model describing a mechanism for the 
formation of integrable turbulence via BG fission of a semiclassical elliptic potential augmented by weak noise.
Our analysis introduces a natural interpretation of a BG as a ``composite SG'', 
consisting of two distinct components: a regular SG plus a soliton condensate.
%
The analytical model comprises a one-parameter family of such BGs, which interpolates between a pure soliton condensate (as $m\to0$) and a rarified SG (as $m\to1$) \cite{Venakides,KMM2003,EJLM,TVZ2004}. 
Intermediate values of $m$ give rise to a mixed regime that interpolates between the above two extremes.  
We validated the theoretical results with direct numerical simulations of the BG fission and demonstrated the establishment, at large $t$ of a statistically  stationary integrable turbulence field characterised by the kurtosis value which is shown analytically and confirmed numerically to be twice as large as the fourth normalized moment of the initial elliptic potential, implying, in particular,  the presence of rogue waves for all nonzero values of $m$.   

We expect that similar results will hold for arbitrary real single-lobe periodic potentials.
Our results also open up a number of interesting avenues for further research. 
Since the ZS problem is common to all equations of the AKNS hierarchy, the results of this Letter will generate SGs/BGs for all such equations,
which although spectrally equivalent, exhibit qualitatively different dynamics.
The study of their resulting SGs/BGs is threfore an interesting open question. 
For example, 
despite some recent work \cite{Pelinovsky2016,Girotti2023}, 
the spectral theory of SGs for the focusing modified Korteweg-deVries (mKdV) equation is still open,
and we expect that the mKdV soliton gas phenomenology will be different than that of both the KdV and focusing NLS gases, potentially involving solitons of both polarities in the same gas. 

\paragraph{Acknowledgement}
We acknowledge fruitful discussions at the Dispersive Hydrodynamics Program hosted by the Issac Newton Institute for Mathematical Sciences in 2022.
We also thank Mark Hoefer for useful suggestions.  
The work of G.B.\ and A.T.\ was partially supported by the National Science Foundation under grants DMS-2009487 and DMS-2009647, respectively, and that of X.D.L.\ by the National Natural Science Foundation of China under grant number 12101590.

\section*{Appendix}
\label{s:supplement}
\def\theequation{A\arabic{equation}}
\setcounter{equation}0
\setcounter{figure}0
\def\thefigure{A\arabic{figure}}

\noindent 
In this appendix we provide additional details on the results presented in the main text. 

\medskip
\paragraph{Jacobi dn potential and the Lax spectrum}
In \cite{BLOT2023} we studied the focusing NLS equation
\be
\label{e:suppnls}
iq_t + q_{xx} + 2|q|^2q = 0,
\ee
where the initial condition (IC) is a multiple Jacobi ``dn'' elliptic function \cite{NIST}:
\be
\label{e:suppICA}
q(x,0) = A\dn(x;m), \quad A\in\R.
\ee
As $m$ goes from 0 to 1 the fNLS dynamics corresponding to $N\dn(x;m)$ initial data interpolates between plane-wave background ($m=0$), to genus $2N-1$ finite-gap solution ($0<m<1$), to pure $N$-soliton ($m=1$). 

To connect the above problem to the semiclassical setting, it is sufficient to rescale $q$ and $t$ to make the initial data independent of $A$, by letting  $q(x,t)\mapsto Aq(x,At)$,
which yields 
\begin{gather} 
\label{e:suppnls2}
i\epsilon q_t + \epsilon^2q_{xx} + 2|q|^2q = 0, \\
\label{e:dn} 
q(x,0) = \dn(x;m),
\end{gather}
with $\epsilon = 1/A$. Hence, studying the semiclassical limit ($\epsilon\to 0$) of Eq.~\eqref{e:suppnls2} with IC~\eqref{e:dn} is equivalent to studying the large-$A$ limit of Eqs.~\eqref{e:suppnls}--\eqref{e:suppICA}. 

\begin{figure}[t!]
\medskip
\centerline{\includegraphics[width=0.24\textwidth]{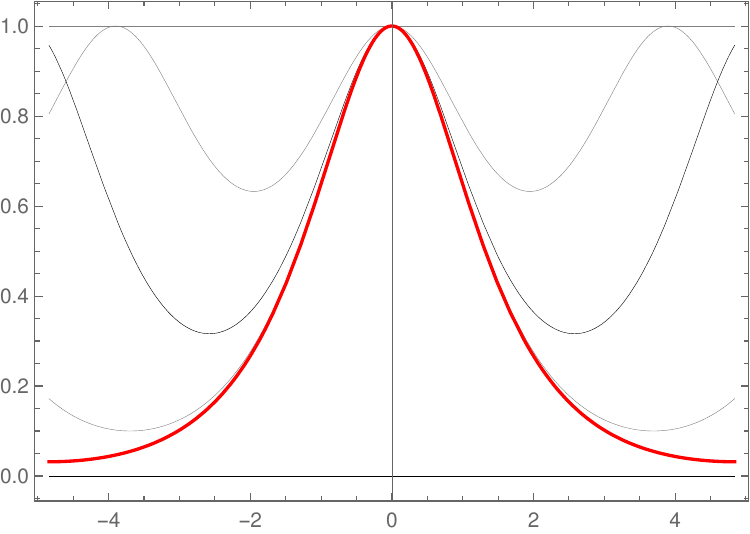}~~
\raise1.2ex\hbox{\includegraphics[width=0.22\textwidth,trim=55 35 40 40,clip]{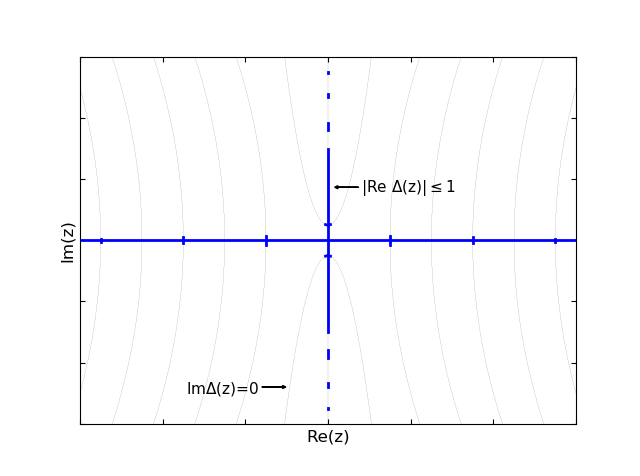}}}
\caption{Left: $\dn(x;m)$ (vertical axis) as a function of $x$ (horizontal axis) for various values of the elliptic parameter $m$.  
From top to bottom: $m=0$ (constant), 0.6 (gray), 0.9 (black), 0.99 (gray) and 0.999 (red).
Right: Diagram illustrating the curves $\Im\Delta(z)=0$ (black) and the Lax spectrum $\Sigma(\L)$ (blue) 
in the complex $z$ plane for the potential~\eqref{e:dn} with generic (non-integer) value of $\epsilon$.}
\kern-1ex
\label{f:a0}
\bigskip
\centerline{\includegraphics[width=0.25\textwidth]{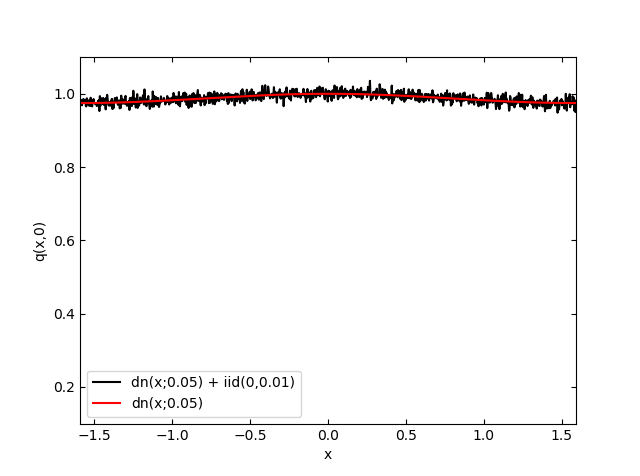}\kern-1em
\includegraphics[width=0.25\textwidth]{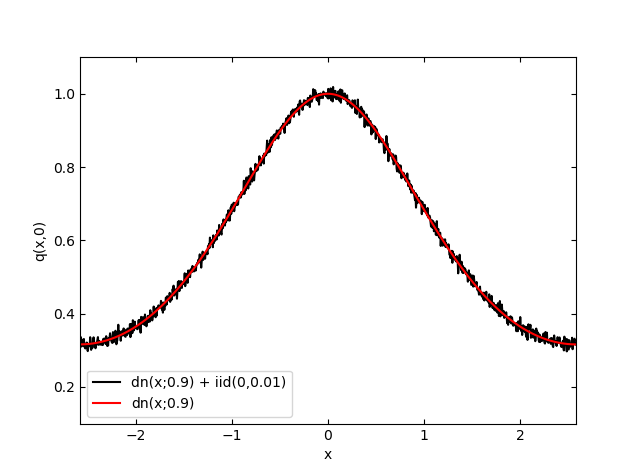}}
\caption{Red curve: Initial condition~\eqref{e:dn}.
Black curve: Same augmented by Gaussian noise with mean $\mu=0$ and standard deviation $\sigma=10^{-2}$.
Left: $m=0.05$. Right: $m=0.9$.}
\label{f:data_m0pt05}
\end{figure}

\begin{figure}[t!]
\centerline{\includegraphics[width=0.25\textwidth]{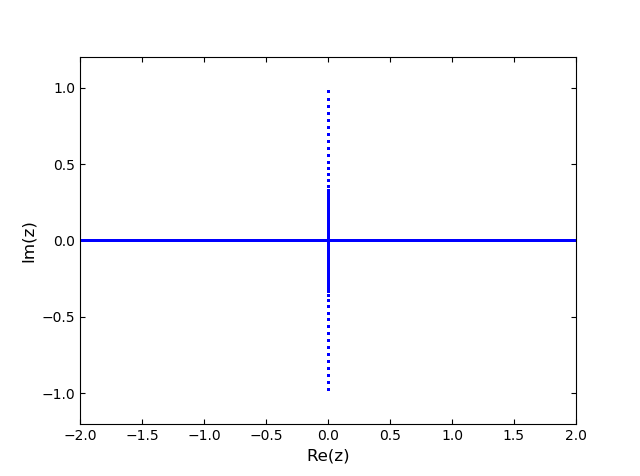}\kern-1em
\includegraphics[width=0.25\textwidth]{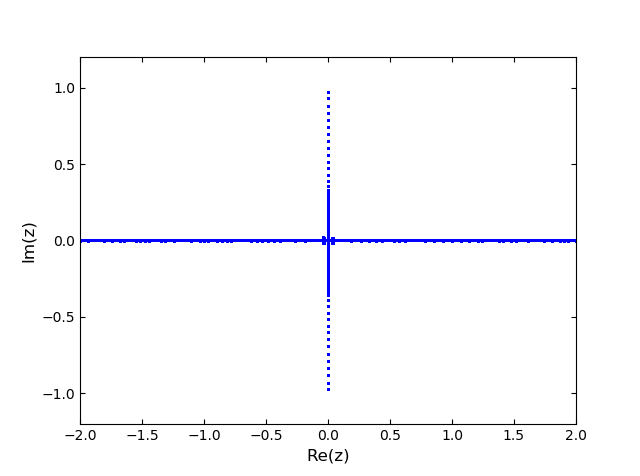}}
\caption{Left: Numerically computed spectrum of the scattering problem for the dn potential with $m=0.9$
via the Fourier-Hill method \cite{DeconinckKutz}. 
Right: Same but with noise added to the potential with $\mu=0$ and $\sigma=10^{-2}$.}
\label{f:noiseaugmentedLaxspectrum}
\end{figure}

Figure~\ref{f:a0}(left) depicts the potential \eqref{e:dn}
for various values of the elliptic parameter $m\in(0,1)$. 
Recall that the real period of the dn potential is $2K_m$, where $K_m=K(m)$ is the complete elliptic integral of the first kind \cite{NIST}. 
Figure~\ref{f:a0}(right) shows the contours $\{z:\Im\D(z)=0\}$ (thin black curves) together with the Lax spectrum (thick blue curves) for a generic (non-integer) value of the semiclassical parameter $\epsilon$.
Figure~\ref{f:data_m0pt05} shows the noise-augmented initial conditions (ICs) for two different values of $m$ and 
Fig.~\ref{f:noiseaugmentedLaxspectrum} compares the Lax spectrum with and without noise, 
demonstrating that the noise does not affect it appreciably. In particular, the number and locations of the solitonic excitations remains unchanged.

In \cite{BLOT2023} we proved that the Lax spectrum of the focusing ZS scattering problem associated
with Eq.~\eqref{e:suppnls} and dnoidal potential \eqref{e:suppICA} and $A\in\Z$ is comprised of 
$2A$ Schwarz-symmetric bands along the imaginary axis of the spectral variable. 
As a result, the corresponding solutions of focusing NLS are special ``finite-gap'' solutions of genus $2A-1$. Next, we provide some details on how the spectrum can be efficiently computed numerically.

\paragraph{PT-symmetric periodic Schr\"odinger potentials with real spectrum}
When the potential is real-valued as with Eq.~\eqref{e:dn},
$\D(z)$ also possesses an additional symmetry: $\D(-z^*) = \D(z)$, and thus the eigenvalues come in quartets, i.e., $\{z,z^*,-z,-z^*\}$. 
Moreover, the transformation $\@w = \Lambda \@v$, where 
$\@w(x, \lambda)=(w_+,w_-)^T$ and 
\vspace*{-0.6ex}
\be
\Lambda = \frac{1}{\sqrt{2}}\begin{pmatrix*}[r] 1 & i \\ 1 & -i \end{pmatrix*}\,,
\ee
maps the ZS problem into the time-independent Schr\"odinger equation with a complex periodic potential, namely, 
\bse
\be
H_\pm w_\pm = \lambda w_\pm\,,\quad
\lambda = z^2\,,
\ee
where 
\be
H_\pm=-\epsilon^2\partial_x^2 + V_\pm(x)\,,\quad 
V_\pm(x)=\mp i\epsilon q_{x}-q^2\,.
\ee
\ese
Recall that the Schr\"odinger equation with a periodic potential is referred to as Hill's equation \cite{MW1966}.
If $q$ is real and even, then the potentials are PT-symmetric: $V_\pm(-x) = V_\pm^*(x)$.
In particular, for the elliptic potential Eq.~\eqref{e:dn},
$H_{\pm}$ is
\be
H_{\pm} = - \epsilon^{2} \partial_x^2 - \dn^2(x) \pm i\epsilon m\sn(x)\cn(x)\,.
\label{e:reduction1}
\ee
Since $\Sigma(H_+)=\Sigma(H_-)$ it is enough to consider only~$H_-$. 
Below we discuss that, even though $H_\pm$ are not Hermitian, all of their eigenvalues are real.

A further simplification is obtained via the 
change of variable $y= 2\am(x;m)$ [where $\am$ is the Jacobi amplitude], 
which maps Eq.~\eqref{e:reduction1} into a 
complex perturbation of Ince's equation~\cite{MW1966}:
\vspace*{-0.4ex}
\begin{multline}
4\epsilon^{2}[1-m\sin^2(y/2)]\,w_{yy} - \epsilon^2 (m \sin y)\,w_{y} 
\\ 
+ [\lambda + (1-m\sin^2(y/2)) + i\epsilon{\textstyle\frac{m}{2}}\,\sin y]\,w = 0\,.
\label{e:trigonometricODE}
\end{multline}
Bloch-Floquet theory implies that any bounded solutions of Eq.~\eqref{e:trigonometricODE} can be written as $w(y) = \e^{i\nu y} p(y)$,
where $\nu\in\Real$ is the Floquet exponent
and $p(y+2\pi)=p(y)$, and can therefore be expanded in Fourier series, 
with Fourier coefficients given by a three-term recurrence relation.
The eigenvalues of the ZS problem are then related to the (real or) complex values of $\lambda$ for which the above ODE admits bounded solutions. Integer and half-integer values of $\nu$ yield respectively periodic and antiperiodic eigenfunctions.

\paragraph{Three-term recurrence relation}
Any solution of Eq.~\eqref{e:trigonometricODE} that is bounded for all $y\in\R$ can be represented by a Fourier series as
\be
w(y;\lambda) = \e^{i\nu y}\sum_{n\in \Z}c_{n}\e^{i ny}\,,
\label{e:fourier}
\ee
with $\nu\in\R$.
The coefficients $\{c_n\}_{n\in\Z}$ are given by the three-term recurrence relation
\be
\label{e:Fourierrecurrence}
\alpha_{n}c_{n-1} + (\beta_{n}-\lambda)c_{n} + \gamma_{n}c_{n+1} = 0\,,
\ee
where
\bse 
\label{e:alphabetagammadef}
\begin{gather}
\alpha_{n} = - m\,[\half-\epsilon(n+\nu-1)][\half + \epsilon (n+\nu-\half)]\,, \\
\beta_{n} = (1-{m}/{2})[\epsilon^{2}(2n+2\nu)^2 - 1]\,, \\
\gamma_{n} = - m\,[\half - \epsilon (n+\nu+1)][\half + \epsilon (n+\nu+\half)]\,,
\end{gather}
\ese
$n\in\Z$. 

\paragraph{Elliptic finite-band potentials}
When $\epsilon=1/N$, some of the coefficents in Eqs.~\eqref{e:alphabetagammadef} vanish, and as a result it is possible to decompose the doubly-infinite recurrence relation into two semi-infinite ones.
Specifically, let 
\bse
\label{e:B_oinftypm}
\begin{align}
B_o^\pm &= \begin{pmatrix} \beta_j & \gamma_j  \\ \alpha_{j+1} & \beta_{j+1} & \gamma_{j+1}  \\  & \ddots & \ddots & \ddots \end{pmatrix},
\\
B_\infty^\pm &= 
\begin{pmatrix} \beta_{-j} &  \alpha_{-j} \\ \gamma_{-j-1} & \beta_{-j-1} & \alpha_{-j-1} \\  & \ddots & \ddots & \ddots 
\end{pmatrix},
\end{align}
\ese
with  $\nu = N/2$ and $j=0$ for the minus sign and $\nu = (1 - N)/2$ and $j=1$ for the plus sign.
The periodic and antiperiodic spectrum of $H_-$ is the union of the spectra of $B_o^\pm$ and $B_\infty^\pm$.
Specifically: 
when $N$ is even, $B_o^-$ and $B_\infty^-$ yield the periodic eigenvalues and $B_o^+$ and $B_\infty^+$ the antiperiodic ones,
and viceversa when $N$ is odd.
In \cite{BLOT2023} we proved that the eigenvalues of all four of these half-infinite matrices are real.
This is the key to prove that, for any  
$m\in(0,1)$, the potential $q$ in Eq.~\eqref{e:dn} for 
the focusing ZS scattering problem is finite-band if and only if $\epsilon = 1/N$ with $N\in\Natural$. (The result is easily extended to $N\in\Z$ by phase invariance.)
Moreover, if $\epsilon = 1/N$, $q$ is a $2N$-band potential, and
\be 
\Sigma(\L)= \R \cup \Big(\!\!\bigcup_{n=1}^{N} [-i\eta_{2n},-i\eta_{2n-1}]\cup [i\eta_{2n-1},i\eta_{2n}]\Big)\,,
\ee 
where $0<\eta_1<\eta_2<\cdots<\eta_{2N}<1$.
This implies that the spectral curve and the flow induced by each member of the Ablowitz-Kaup-Newell-Segur (AKNS) hierarchy has finite genus $2N-1$.   
Importantly, the finite truncations of the matrices~\eqref{e:B_oinftypm} provide an efficient way to numerically compute the spectrum.

\paragraph{Semiclassical WKB analysis of the ZS problem}
Next we consider the semiclassical limit of the spectrum, namely the limiting behavior of bands and gaps as $\epsilon\to0$.
It is useful to briefly recall the asymptotic analysis of the focusing ZS scattering problem via WKB methods
from~\cite{BO2020}. 
We emphasize that the results of this and the next few sections apply to a broad class of potentials, not just dn.

\begin{figure}[t!]
\centerline{\includegraphics[width=0.255\textwidth]{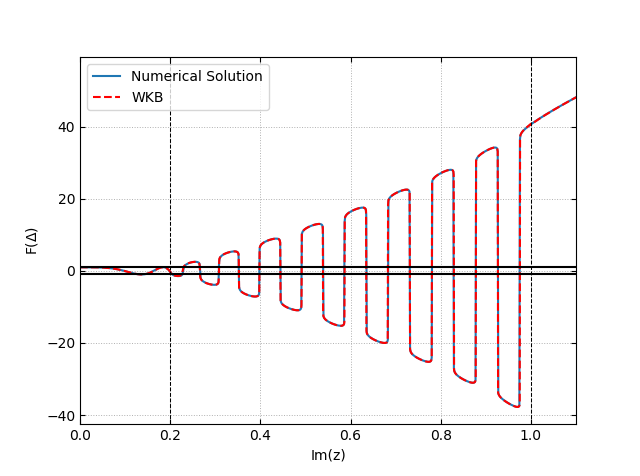}\kern-1em
\includegraphics[width=0.255\textwidth]{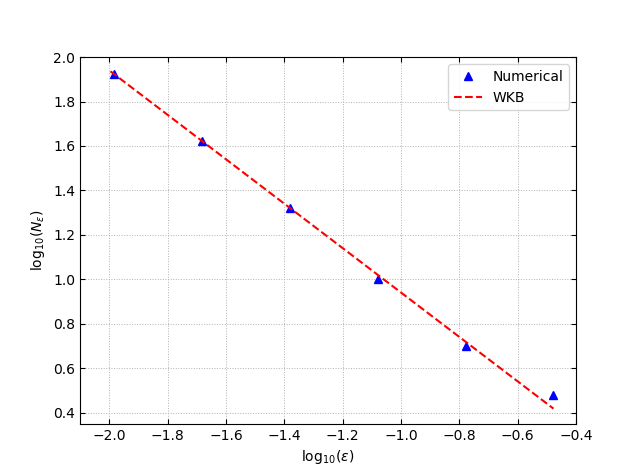}
}
\caption{Left: the numerically computed Floquet discriminant (in blue) for the potential~\eqref{e:dn} with $m=0.96$ and $\epsilon=1/20$, using fourth-order Runge-Kutta versus the WKB approximation (in red). 
The black solid horizontal lines show $\D=\pm 1$. 
The black dashed vertical lines show $\Im(z)=q_{\rm min}$ and $\Im(z)=q_{\rm max}$. 
Note that, to scale the plot, the function $F(\tau)= \tau$ for $|\tau|\le 1$ and $F(\tau)={\rm sgn}(\tau)(1+\log_{10}|\tau|)$ for $|\tau|>1$ is plotted.
Right: The number of spectral bands for $z\in\Gamma_g$ as predicted by WKB (red curve) versus the numerically computed
values (blue triangles) as a function of~$\epsilon$.}
\kern-1ex
\label{f:a1}
\end{figure}

Suppose that $q(x)$ is the $2L$-periodic extension of a real, even, non-negative single-lobe potential.
Thus, $q$ has one maximum and one minimum in $(-L,L]$, which without loss of generality can be taken to be respectively at $x=0$ and $x=L$. 
To avoid trivial cases, assume that $q$ is not constant.
Let
\bse
\be
q_{\rm max} = q(0) \,,\quad
q_{\rm min} = q(L)\,,
\ee
and 
\be
\label{e:s-int}
s(x,z) = \int_{-x_o(z)}^{x}\sqrt{|q^2(u)+z^2|}\,\d u\,,
\ee
\ese
where $z\in\R\cup i\R$, and $x_o(z)$ is a simple (real) turning point. 
Without loss of generality we can limit ourselves to considering $\Im z\ge0$ thanks to the Schwarz symmetry of the spectrum.
Then, as $\epsilon\to 0$ (see \cite{BO2020,BOT}):
\begin{itemize}
\item[(i)] 
For $z\in \Gamma_o\cup\R$, where $\Gamma_o = (0,iq_{\rm min})$, one has $q^2(x)+z^2>0$ for $x\in(-L,L)$. 
There are no turning points, and
\be
\D(z;\epsilon) \sim \cos(s_{\rm i}(z)/\epsilon)\,,
\label{e:cos}
\ee
where $s_{\rm i}(z) = \int_{-L}^{L}\sqrt{q^2(x)+z^2}\,\d x$.
This range corresponds to the region in white in Fig.~\ref{f:spectrum}.
\item[(ii)] 
For $z\in \Gamma_g$, where $\Gamma_g = (iq_{\rm min},iq_{\rm max})$, there are two real symmetric turning points $\pm x_o$, i.e., values at which $q^2(\pm x_o)+z^2=0$. In this region (see Fig.~\ref{f:a1})
\be
\D(z;\epsilon) \sim \cos(s_1(z)/\epsilon)\cosh(s_{2,\epsilon}(z)/\epsilon)\,,
\label{e:coscosh}
\ee
where $s_1(z)=s(x_o,z)$ and $s_{2,\epsilon}(z)=\epsilon\ln 2-2s(-L,z)$.
This range corresponds to the region in gray in Fig.~\ref{f:spectrum}.
\item[(iii)] 
For $z\in (iq_{\rm max},i\infty)$ one has $q^2(x)+z^2<0$ for $x\in (-L,L)$. There are no turning points and
\be
\D(z;\epsilon) \sim \cosh(s_{\rm ii}(z)/\epsilon)\,,
\label{e:cosh}
\ee
where $s_{\rm ii}(z)=\int_{-L}^{L}\sqrt{-q^2(x)-z^2}\,\d x$. 
\end{itemize}
Figure~\ref{f:a1} shows a comparison between the predicted behavior of $\Delta(z,\epsilon)$ and the numerically computed value.

\paragraph{Semiclassical limit of bands and gaps in~$\boldsymbol{\Gamma}_g$}
When $z\in(iq_{\rm min}, iq_{\rm max})$,
Eq.~\eqref{e:coscosh} implies that the bands are approximately centered at the roots $\{z_n\}$ of the equation $s_1(z_n)=(n-\half)\pi\epsilon$. Thus, the asymptotic number of bands for $z\in(iq_{\rm min},iq_{\rm max})$ is 
\bse
\be
N^\epsilon_\mathrm{bands} \sim \Big\lfloor \frac{J_g}{\pi\epsilon} \Big\rfloor, \qquad \epsilon\to 0\,,
\label{e:nbands}
\ee
where 
\be
J_g = s_1(iq_{\rm min}) = \int_{-L}^{L}\sqrt{q^2(x) - q^2_{\rm min}}\,\d x\,.
\label{e:Jgdef}
\ee
\ese
and $\lfloor\cdot\rfloor$ is the floor function.

Next, consider the limit $\epsilon\to 0$ and $n\to\infty$ satisfying $0<n\epsilon<J_g/\pi$ and $n$ is the band index.
Denote the $n$-th band width as $w_n^{\epsilon} = |z_n^{+}-z_n^{-}|$, where 
$\Delta(z_n^{\pm}) = \pm 1$, respectively.
Similarly, denote the $n$-th gap width as $g_n^{\epsilon}= |z_{n+1}^{\pm}-z_n^{\pm}|$.
Using $\D(z_n^{\pm})=\pm 1$ together with Eq.~\eqref{e:coscosh}, one gets 
\be
\label{e:asymbeta}
w_n^{\epsilon} \sim  \frac{4\epsilon}{|s_1'(z_n)|}\e^{-s_{2,\epsilon}(z_n)/\epsilon},\qquad \epsilon\to 0\,.
\ee
Moreover, since $g_n^{\epsilon} \sim |z_{n+1}-z_{n}|$ and $s_1(z_{n+1})-s_1(z_n) = \pi\epsilon$ it follows that
\be
\label{e:asympgamma}
g_n^{\epsilon} \sim \frac{\pi\epsilon}{|s_1'(\xi_n)|},\qquad \epsilon\to 0\,,
\ee
where $\Im (z_{n+1}) < \Im(\xi_n) < \Im (z_n)$. Finally, the band-to-gap ratio $w_n^{\epsilon}/g_n^{\epsilon}\to 0$ as $\epsilon\to 0$ exponentially fast in this region.

\paragraph{Effective solitons in the fNLS equation}
Recall that, for localized potentials (i.e., $q(x,t)\to0$ rapidly as $|x|\to\infty$), solitons are parameterized by the discrete eigenvalues of the ZS scattering problem. 
In the periodic problem, nonlinear excitations are considered to be ``effective solitons'' if the relative band width is less than some small threshold parameter
\cite{BO2020,Osborne,Deng2016},
i.e., if $W_n < \kappa$, where $0<\kappa\ll 1$ and
\be
\label{e:relbandwidth}
W_n^{\epsilon} = \frac{w_n^{\epsilon}}{w_n^{\epsilon}+g_n^{\epsilon}}\,.
\ee 
Thus, rewriting the relative band width as
\be
W_n^{\epsilon} = \frac{w_n^{\epsilon}}{g_n^{\epsilon}}\Big(\frac{1}{1+w_n^{\epsilon}/g_n^{\epsilon}}\Big)\,,
\ee
one can show that 
\be
W_n^{\epsilon} \sim \frac{4}{\pi}\e^{-s_{2,\epsilon}(z_n)/\epsilon},\qquad \epsilon\to 0\,.
\ee
Note that for $z\in (iq_{\rm min},iq_{\rm max})$ the relative band width $W_n$ is monotonic decreasing. Thus, the effective solitons are confined to the interval 
\be
z\in (z_s, iq_{\rm max}) \subset (iq_{\rm min}, iq_{\rm max})\,,
\ee
where $z_s$ is the unique solution to $W_n=\kappa$. For $z\in(iq_{\rm min},iq_{\rm max})$ let $s_2(z) = -s(-L,z)\ge 0$ 
(see Eq.~\eqref{e:s-int}), then to leading order $z_s$ is given implicitly by
\be
\label{e:implicit-eq}
s_2(z_s) = \frac{\epsilon}{2}\ln \Big(\frac{2}{\pi\kappa}\Big)\,,
\ee
and the number of effective solitons is given by
\be
\label{e:nsolitons}
N_\mathrm{solitons} = \Big\lfloor \frac{s_1(z_s)}{\pi\epsilon} + \frac{1}{2} \Big\rfloor\,.
\ee 
Expanding $s_2(z)$ about $z=iq_{\rm min}$ and evaluating at $z_s$, for any $\kappa>0$ it follows that $z_s\to iq_{\rm min}$ as $\epsilon\to 0$. 
Thus, in the semiclassical limit the entire interval $(iq_{\rm min},iq_{\rm max})$ is comprised of an infinite ensemble of effective solitons.

\eject
\paragraph{Semiclassical limit of bands and gaps in~$\Gamma_o$}
Recall that, when $q=\dn(x;m)$ with $m\in(0,1)$ and $\epsilon=1/N$ with $N\in\Natural$ there are precisely $N$ bands for $z\in(0,iq_{\max})$. 
Hence, not all bands become effective solitons in the limit $\epsilon\to 0$ (see Fig.~\ref{f:spectrum}).
For $z\in\Gamma_o$, following similar arguments as for $z\in\Gamma_g$, one obtains
\be
w_n^{\epsilon} = O(\epsilon),\qquad \epsilon\to 0\,. 
\ee
Thus, at $z_*=iq_{\min}= i\dn(K_m;m)$, which is the boundary between the intervals 
$\Gamma_g=(iq_{\min},iq_{\max})$  
and $\Gamma_o = (0,iq_{\min})$,  
there is a transition from exponentially decaying band widths (giving rise to a soliton gas), to algebraically decaying band widths (giving rise to a soliton condensate). 
Thus, we have the physically realistic scenario of a \textit{generalized breather gas} --- namely, a soliton gas on a condensate background.

The number of bands with support in $\Gamma_o$ can be computed by noting that 
Eq.~\eqref{e:cos} holds in this interval with exponentially small corrections at $\D(z)=\pm 1$. 
Thus, bands in this region are approximately centered at the roots $\{z_n\}$ of the equation $s_{\rm i}(z_n)=(n-\half)\pi\epsilon$. Thus, the asymptotic number of bands for $z\in(0,iq_{\rm min})$ is 
\bse
\be 
\label{e:nbands2}   
N_{\rm bands}^{\epsilon} \sim \Big\lfloor \frac{J_o}{\pi\epsilon}\Big\rfloor,\qquad 
\epsilon\to 0\,,
\ee 
where 
\be 
J_o = s_{\rm i}(0)-s_{1}(iq_{\rm min}),
\label{e:Jdef}
\ee 
and
\be
s_{\rm i}(0)=\int_{-L}^{L}q(x)\,\d x\,.
\ee
\ese

\paragraph{Spectral characteristics of the composite soliton gas}
%
%
Recall that the density of bands $\varphi(z)$ and scaled logarithmic band width $\nu(z)$ are two key quantities in the spectral theory of soliton gases. 
Following \cite{Wang2022} we have that, for $z\in\Gamma_g$, and the dn potential \eqref{e:dn}
\bse
\begin{align}
\label{e:densitygas}
\varphi(z) &= \frac{2|z|}{J}\int_{0}^{\dn^{-1}(|z|)}\frac{\d x}{\sqrt{\dn^2(x;m)+z^2}}\,, \\
\nu(z) &= \frac{2\pi}{J} \int_{\dn^{-1}(|z|)}^{K_m}\sqrt{|\dn^2(x;m)+z^2|}\,\d x\,,
\end{align}
\ese 
with $J$ is a suitable normalization constant, determined below. 
%
Thus we can easily express the spectral scaling function as
\vspace*{-1ex}
\be
\label{e:specscaling}
\sigma(z) = \frac{2\nu(z)}{\varphi(z)}.
\ee
Similarly, for $z\in\Gamma_o$, we have 
\bse
\begin{align}
\label{e:condensatedensity}
\varphi(z) &= \frac{2|z|}{J}\int_{0}^{K_m}\frac{\d x}{\sqrt{\dn^2(x;m)+z^2}}\,,
\\
\sigma(z) &= 0\,.
\end{align}
\ese
(In $\Gamma_o$, the spectral scaling function $\sigma(z)$ is zero as expected for a condensate.)
Finally, ensuring that the integral of the density of bands
over the whole support $(0,i)$ is one, 
one can then show that~\cite{Wang2022} 
\be
J = J_g + J_o = \int_{-K_m}^{K_m} \dn(x;m)\,\d x = \pi\,.
\ee

\paragraph{Limits $m\to 0$ and $m\to 1$}
Next we discuss the density of bands formulae in the limits $m\to 0^+$, and $m\to 1^-$.

Recall that $\dn(x;0)\equiv 1$ and $K(0)=\pi/2$. Thus, $\q_{\rm min}\to 1$ as $m\to 0^+$ and Eq.~\eqref{e:condensatedensity} has support $z\in(0,i)$. Moreover,
$J_o \to \pi$ as $m\to 0^+$. Thus,
\be
\varphi(z) \to \frac{|z|}{\sqrt{1+z^2}}, \qquad m\to 0^+\,,
\ee
which corresponds to Weyl's distribution as expected (see \cite{PRL2019v123p234102}).

Next, recall that $\dn(x;1)\equiv \sech(x)$ and $K(m)\to\infty$ as $m\to 1^-$. Thus, $q_{\rm min}\to 0$ as $m\to 1^-$ and Eq.~\eqref{e:densitygas} has support $z\in(0,i)$. Moreover, $J_g\to \pi$ as $m\to 1^-$, and $\dn^{-1}(|z|) \to {\rm arcsech}(|z|)$ as $m\to 1^-$. 
Thus,
in this limit we get a uniform distribution:
\be
\varphi(z) \to 1, \qquad m\to 1^-\,.
\ee

\paragraph{Nonlinear dispersion relations}
Together with the set $\Gamma_+ = \Gamma_g\cup\Gamma_o$, 
the spectral scaling function~$\sigma(z)$ 
determines the integral equation (the first nonlinear dispersion relation, NDR) DOS $f(z)$ for soliton gases:
\begin{gather} \label{dr_soliton_gas1}
\int _{\Gamma_+}\log \left|\frac{\zeta-z^*}{\zeta-z}\right|f(\zeta)\,\d|\zeta|+\sigma(z)f(z) = \Im z.
\end{gather}
(Note that in the portion of the spectrum corresponding to a soliton condensate, i.e., $\Gamma_o$, one has $\sigma(z)=0$.)
The second NDR involves the density of fluxes, but this quantity is zero in our case since all the nonlinear excitations have zero velocity.
It was proven in \cite{Wang2022} that $f(z)=r\varphi(z)$ solves the first NDR for soliton gases, where 
\be\label{r-const}
r= \frac{J}{2\pi L} = \frac1{2K_m}\,.
\ee 
As a result, expressions~\eqref{e:densitygas} and~\eqref{e:condensatedensity} also yield the density of states (DOS) $f(z)$ for the dn potential \eqref{e:dn}.

\paragraph{Kurtosis and Jensen's inequality}
In the main text,we have shown that the kurtosis $\kappa$ of a fully developed SG
generated by a deterministic real and even single-lobe initial condition $q(x,0)$ 
is given by $\kappa = 2\kappa_0$, 
where 
\be
\kappa_0 = \frac{L\int_0^L|q(x,0)|^4\,\d x}{\big(\int_0^L|q(x,0)|^2\,\d x\big)^2}\,.
\ee
Next, the rescaling $y = x/L$ and $\~q(y,0) = q(Ly,0)$ implies
\be
\kappa_0 = 
\frac{\int_0^1|\tilde{q}(y,0)|^4\,\d y}{\big(\int_0^1|\tilde{q}(y,0)|^2\,\d y\big)^2}\,.
\ee
Then Jensen's inequality \cite{NIST} implies that $\kappa_0\ge1$, 
with the equality holding only if $q(x,0)$ is constant.

\begin{table}[b!]
\begin{center}
\scriptsize
\begin{tabular}{ |c|c|c|c|c|c| }
\hline
$m$ & $\kappa_o$ & $\kappa_{\infty}$ & ~~$T_\infty$~~ & ~$~M$~~ & ~~$N_\mathrm{Fourier}$~~ \\
\hline 
$0.05$ & 1.007 & 1.92 & $4$ & 500 & 1024 \\
\hline 
$0.3$ & 1.016 & 1.96 & $4.5$ & 400 & 1024 \\
\hline
$0.5$ & 1.059 & 2.07 & $4$ & 400 & 2048 \\
\hline 
$0.7$ & 1.17 & 2.37 & $2.5$ & 200 & 4096 \\
\hline
$0.8$ & 1.29 & 2.66 & $2.5$ & 200 & 4096 \\
\hline 
$0.9$ & 1.53 & 3.022 & $3.0$ & 400 & 4096 \\
\hline 
\end{tabular}
\end{center}
\caption{Numerical simulation parameters and results, where
$N_\mathrm{Fourier}$ is the number of Fourier modes used,
$M$ is the number of simulations (each of which corresponds to an independent noise realization), and 
$T_\infty$ is the numerically observed approximate thermalization time.}
\label{t:values}
\end{table}

\begin{figure}[b!]
\centerline{\includegraphics[width=0.255\textwidth]{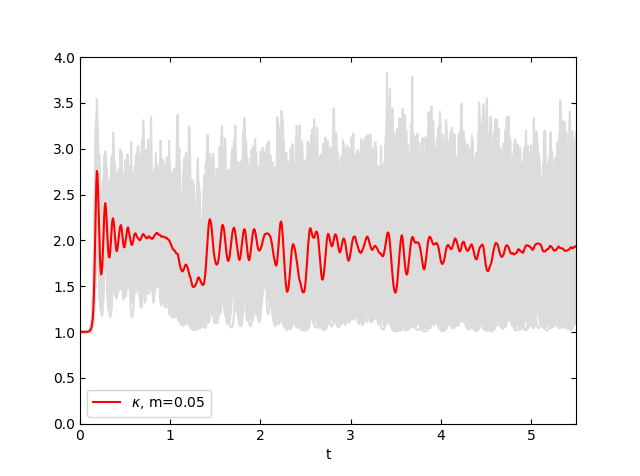}\kern-1em
\includegraphics[width=0.255\textwidth]{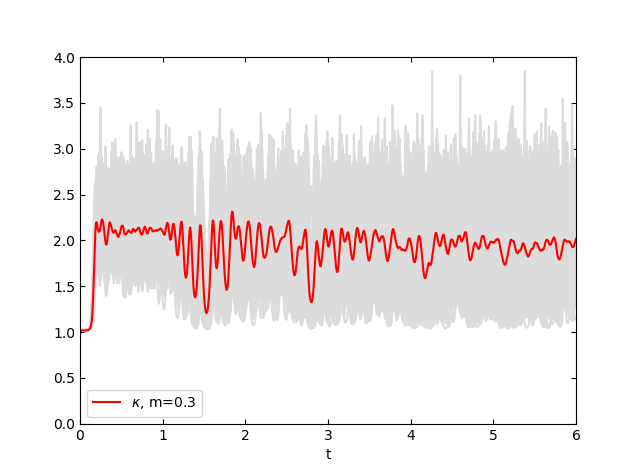}}
\centerline{\includegraphics[width=0.255\textwidth]{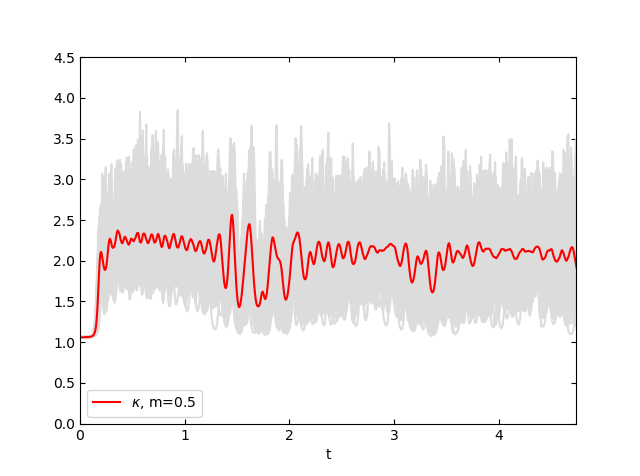}\kern-1em
\includegraphics[width=0.255\textwidth]{k4_simulations_m0pt7.png}}
\centerline{\includegraphics[width=0.255\textwidth]{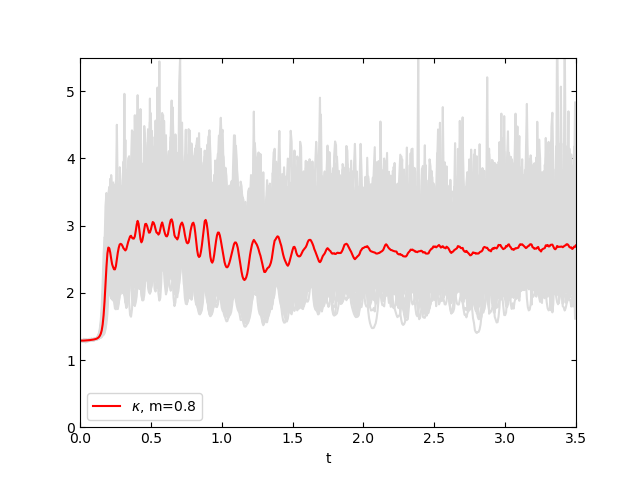}\kern-1em
\includegraphics[width=0.255\textwidth]{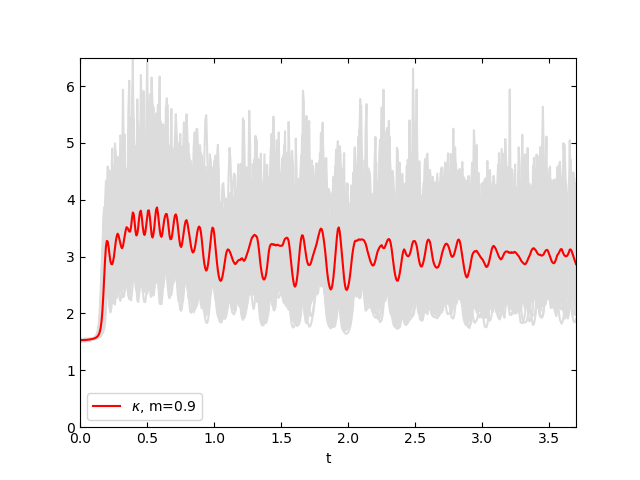}}
\caption{Time evolution of the kurtosis with $\epsilon = 1/20$ and $\sigma=10^{-2}$.
Gray regions: Envelope of 100 ensemble realizations.
Red curves: Ensemble average of $M$ realizations.
Top left: $m=0.05$ ($N=500$).
Top right: $m=0.3$ ($N=400$).
Middle left: $m=0.5$ ($N=400$).
Middle right: $m=0.7$ ($N=200$).
Bottom left: $m=0.8$ ($N=200$).
Bottom right: $m=0.9$ ($N=400$).
}
\label{f:k4_time}
\end{figure}

\begin{figure}[b!]
\centerline{\hglue-1em\includegraphics[trim=35 0 50 30,clip,height=0.205\textwidth]{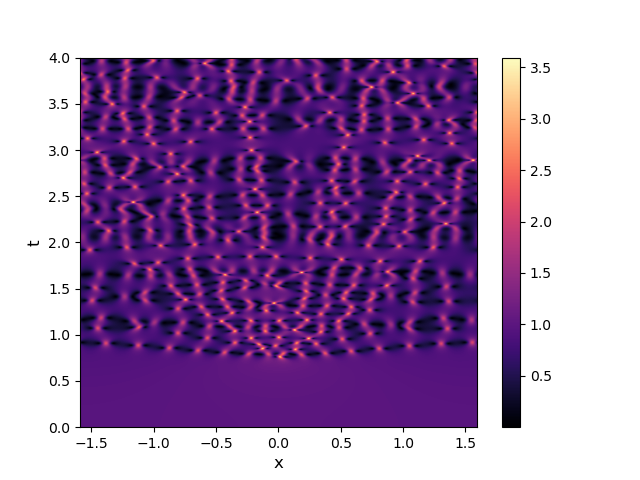}\kern0em
\includegraphics[trim=35 0 50 30,clip,height=0.205\textwidth]{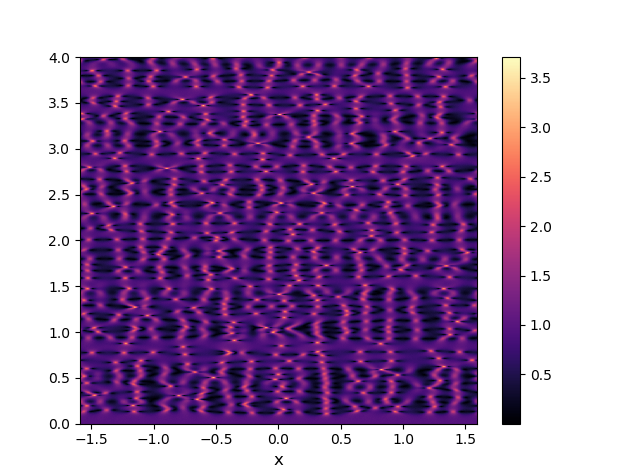}\kern-2em}
\centerline{\includegraphics[width=0.265\textwidth]{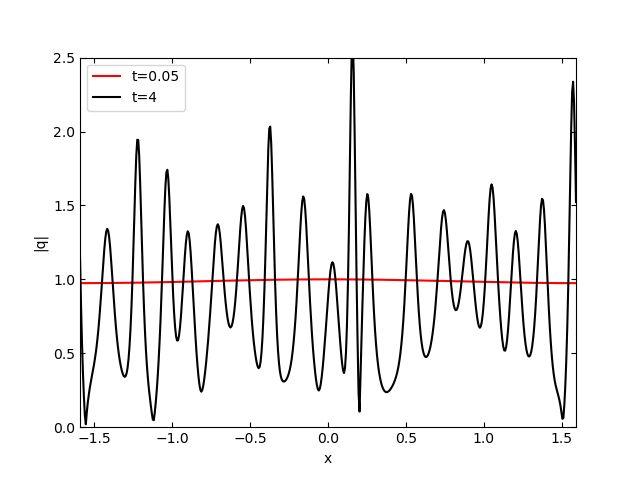}\kern-1.2em
\includegraphics[width=0.265\textwidth]{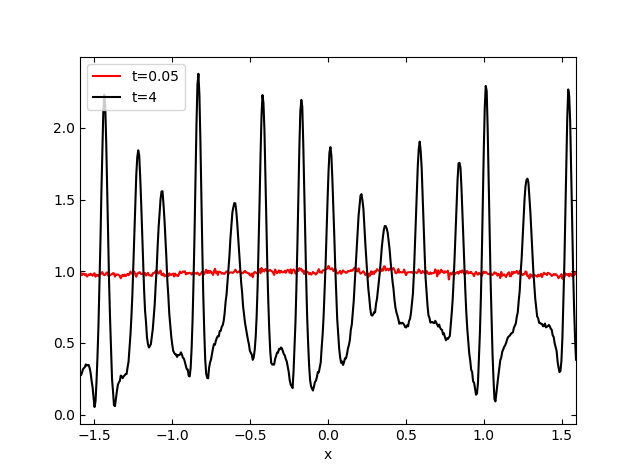}}
\caption{Numerical solution of fNLS with dn IC, with and without noise, and $\epsilon=1/20$. 
Column 1: $\sigma=0$ (without noise);
Column 2: $\sigma=10^{-2}$ (with noise);
Row 1: Density plot, $m=0.05$;
Row 2: Solution at $t=0.05$ and $t=4$ for $m=0.05$.}
\label{f:density_plots}
\end{figure}

\paragraph{Numerical methods}
We performed several numerical experiments related to the kurtosis of randomly generated solutions to the fNLS equation. 
In particular, we studied the time evolution of fNLS in the semiclassical limit with elliptic IC plus a complex perturbation of independent and identically distributed normal random variables with mean $\mu=0$ and standard deviation $\sigma=10^{-2}$.

Due to modulational instability of fNLS the numerical computation of solutions is a delicate topic, especially in the semiclassical limit. To this end, in this work all numerical simulations of the fNLS equation were performed using an eighth-order split-step method \cite{Yoshida}. 
Moreover, to ensure numerical accuracy the isospectral property of the scattering data was confirmed at several data points as the simulated solution evolved in time. In particular, the Lax spectrum was computed using the Floquet-Fourier-Hill method \cite{DeconinckKutz}. 
Finally, to reduce the number of simulations needed a double-averaging technique was used to compute the kurtosis. 
That is, at each time we compute the spatial average of $|q|^{2n}$, $n=1,2$, and then we compute the ensemble average. 

\begin{figure}[t!]
\centerline{\hglue-1em\includegraphics[trim=35 0 50 30,clip,height=0.205\textwidth]{numsoln_nls_dn_m0pt7_eps0pt05.png}\kern0em
\includegraphics[trim=35 0 50 30,clip,height=0.205\textwidth]{numsoln_nls_dn_m0pt7_eps0pt05_noise}\kern-2em}
\centerline{\includegraphics[width=0.265\textwidth]{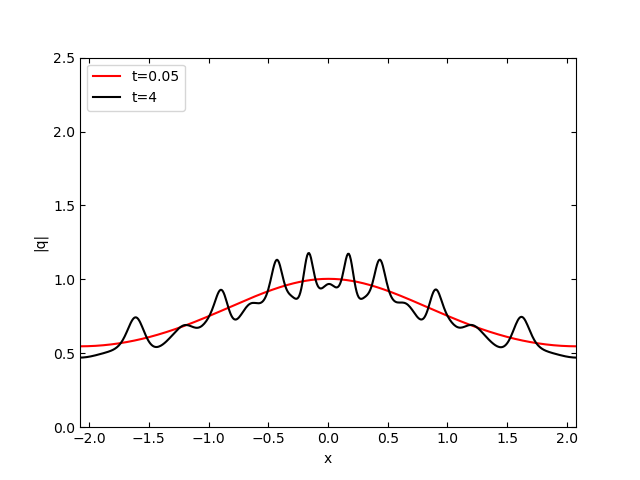}\kern-1.2em
\includegraphics[width=0.265\textwidth]{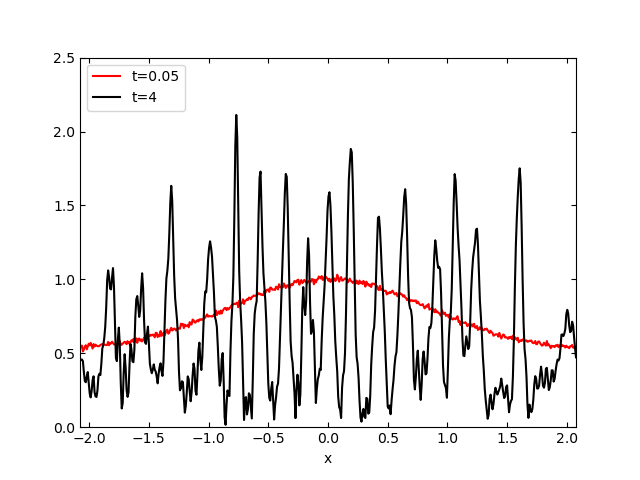}}
\caption{Same as Fig.~\ref{f:density_plots}, but for 
$m=0.7$.}
\label{f:density_plots2}
\bigskip
\centerline{\hglue-1em\includegraphics[trim=35 0 50 30,clip,height=0.205\textwidth]{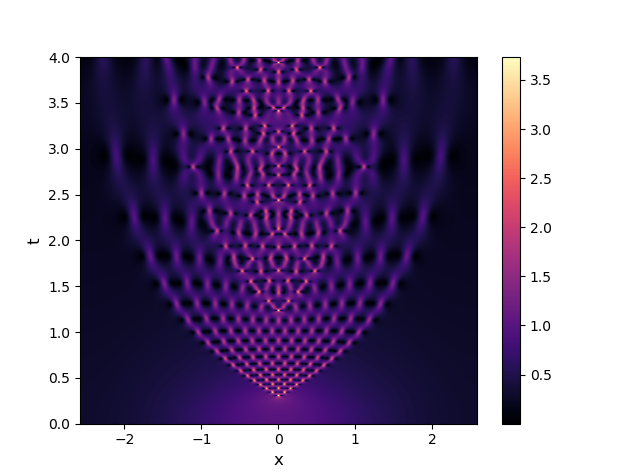}\kern0em
\includegraphics[trim=35 0 50 30,clip,height=0.205\textwidth]{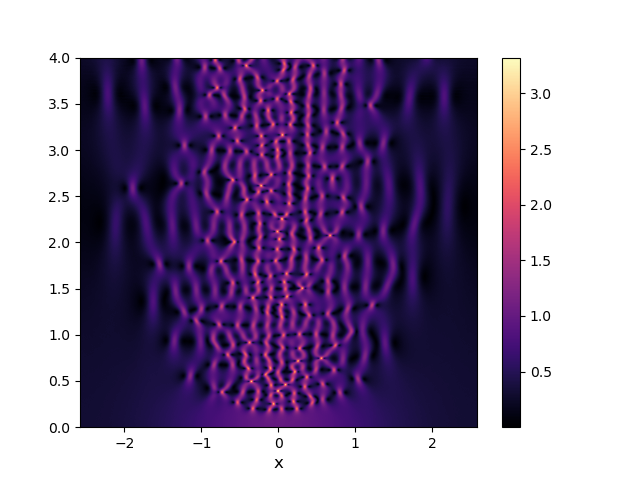}\kern-2em}
\centerline{\includegraphics[width=0.265\textwidth]{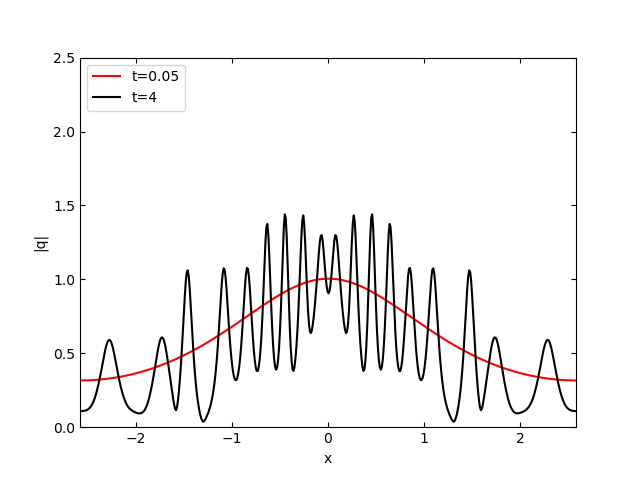}\kern-1.2em
\includegraphics[width=0.265\textwidth]{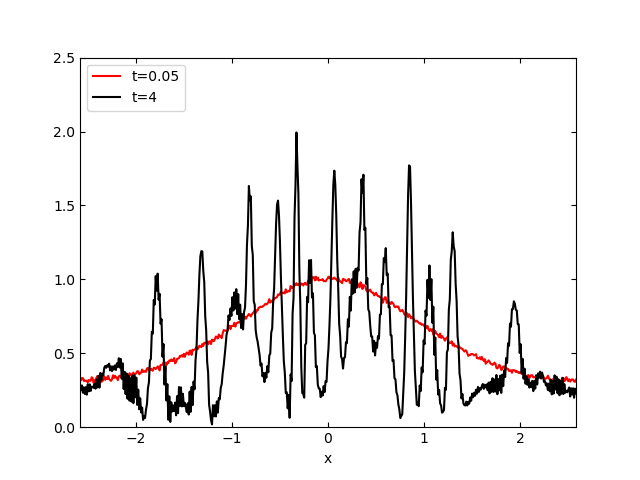}}
\caption{Same as Fig.~\ref{f:density_plots}, but for 
$m=0.9$.}
\label{f:density_plots3}
\end{figure}

\paragraph{Further simulation results}
For consistency, the numerical experiments presented in this work related to the kurtosis have $\epsilon=1/20$, and $\sigma=10^{-2}$ fixed for all values of $m\in(0,1)$ considered. Table~\ref{t:values} provides key parameters used in the simulations as well as some of the key quantities related to the kurtosis experiments. Moreover, Fig.~\ref{f:k4_time} illustrates the kurtosis for $m=0.05$, $0.3$, $0.5$, $0.7$, $0.8$, and $0.9$ as a function of time. The gray region is the envelope of 100 ensemble realizations. The red curves are ensemble averages of $M$ realizations. Notice for each $m$ the ensemble average of the kurtosis settles at large times (see Table~\ref{t:values} for approximate thermalization times).

To further illustrate thermalization of the SG at large times several plots of the numerically computed solution of fNLS in the semiclassical limit are provided. In each case, the left column depicts the time evolution of the dn potential \eqref{e:dn} while the right column depicts the time evolution of the dn potential \eqref{e:dn} with a small complex-valued random perturbation. Further, the top row depicts a density plot while the bottom row depicts the solution at a particular time. In Fig~\ref{f:density_plots} we have $m=0.05$ which is a small sinusoidal perturbation of the constant background studied in \cite{PRE98p042210}. In Fig.~\ref{f:density_plots2} $m=0.7$, and in Fig.~\ref{f:density_plots3} $m=0.9$. Notice the numerically computed solutions with noise are spatially homogeneous at large times.

\def\doibase{http://dx.doi.org/}
\def\reftitle#1,{\relax}
\def\booktitle#1{\textit{#1}}


\begin{thebibliography}{12}
\bibitem{Laxspecdef}
The Lax spectrum $\Sigma(\L)$ is defined as the set of values $z\in\Complex$ such that the focusing ZS scattering problem admits solutions $\@v(x,z)$ that are bounded for all $x\in\Real$.
\bibitem{omit_time_dependence}
Since $\Sigma(\L)$ is invariant when the potential $\q(x,t)$ evolves in time according to the fNLS Eq.~\eqref{e:nls},
we will often omit the time dependence from all quantities below.
\bibitem{reducedDOS}
Note that $f(\eta)$ is a ``reduced'' DOS. 
The ``full'' DOS for the bound state potential is given by $\~f(z)= f(\eta)\delta(\xi)$, where $\xi = \Re z$ 
and $\delta(x)$ is the Dirac delta-function, see \cite{Congy2024}.
\end{thebibliography}
\end{document}